\documentclass[twocolumn,showpacs,preprintnumbers,amsmath,amssymb]{revtex4}

\usepackage{graphicx}
\usepackage{dcolumn}
\usepackage{bm}

\def\be{\begin{equation}}
\def\ee{\end{equation}}
\def\bea{\begin{eqnarray}}
\def\eea{\end{eqnarray}}
\def\NO{\nonumber}
\def\gev{\mathrm{~GeV}}

\begin{document}


\title{Reconciling charmonium production and polarization data within the nonrelativistic QCD framework}


\author{Zhan Sun$^{1,2}$}
\author{Hong-Fei Zhang$^{1}$}
\affiliation{
$^{1}$ Department of Physics, School of Biomedical Engineering, Third Military Medical University, Chongqing 400038, China. \\
$^{2}$ Department of Physics, Chongqing University, Chongqing 401331, China.
}%
\date{\today}

\begin{abstract}
A thorough study reveals that the only key parameter for $\psi$ ($J/\psi$, $\psi'$) polarization is the ratio $\langle O^\psi(^3S_1^{[8]})\rangle/\langle O^\psi(^3P_0^{[8]})\rangle$,
if the velocity scaling rule holds.
The ordinary fitting precedure is incapable of the determination of this parameter.
We provide a universal approach to fixing the long-distance matrix elements (LDMEs) for the $J/\psi$ and $\psi'$ production.
Further, with the existing data, we implement this approach, and obtain a favorable set of the LDMEs,
and manage to reconcile the charmonia production and polarization experiment.
A quantitative analysis indicates that theoretical corrections to the short-distance coefficients change only the values of the LDMEs but not the phenomenological results.
\end{abstract}

\maketitle

{\it Introduction.}---
Nonrelativistic QCD (NRQCD)~\cite{Bodwin:1994jh} is one of the most successful effective theory describing quarkonium productions and decays (as a review, see e.g.~\cite{Brambilla:2010cs}).
Despite its contributions, still, it is facing challenges from many aspects.
Three groups~\cite{Butenschoen:2012px, Chao:2012iv, Gong:2012ug} succeeded to accomplish QCD next-to-leading order (NLO) calculations of $J/\psi$ production
and polarization at hadron colliders, however, with different fitting strategies, obtained quite different values of the long-distance matrix elements (LDMEs),
consequently leading to different perspective of the polarization puzzle.
Recently, the LHCb Collaboration released their results of $\eta_c$ hadroproduction measurement~\cite{Aaij:2014bga}.
Three groups~\cite{Butenschoen:2014dra, Han:2014jya, Zhang:2014ybe} looked into the experimental data from relatively different angles of view.
Many of the existing works~\cite{Chao:2012iv, Bodwin:2014gia, Faccioli:2014cqa} interpreted the almost unpolarized experimental results of the $J/\psi$ hadroproduction measurements as
the indication of $^1S_0^{[8]}$ dominance, which violates the heavy quark spin symmetry regarding the data in Ref.~\cite{Aaij:2014bga}.
Others~\cite{Butenschoen:2011yh, Butenschoen:2010rq, Gong:2012ug}, even with different philosophy, also came to the similar conclusions.
Refs.~\cite{Han:2014jya, Zhang:2014ybe} remedied the discrepancy between the measurements of $J/\psi$ and $\eta_c$ hadroproduction.
Notably, their LDMEs are consistent with the velocity scaling rule (VSR), which is essential to the NRQCD expansion.
Even so, both of them failed in the explanation of the $J/\psi$ polarization data in midrapidity regions.
The polarization parameter $\lambda$ converges in the ranges
$0.05<\lambda<0.2$ (for $|y|<0.6$, denoted as E1) and $0<\lambda<0.1$ (for $0.6<|y|<1.2$, denoted as E2) for the CMS experiment~\cite{Chatrchyan:2013cla},
and $-0.2<\lambda<0$ for the CDF experiment~\cite{Abulencia:2007us} at $\sqrt{s}=1.96$TeV (denoted as E3).
However, the theoretical predictions~\cite{Han:2014jya, Zhang:2014ybe} of $\lambda$ for E1, E2 and E3 reach 0.4, 0.4 and 0.2, respectively.
Taking E2 as an example, the experimental and theoretical values of the ratio of the transverse cross section to the longitudinal one (denoted as $\xi$) are about 1.2 and 2.3, respectively.
A successful effective theory cannot tolerate so large a discrepancy.
Accordingly, Refs.~\cite{Butenschoen:2014dra, Han:2014jya} both agreed that,
despite that the yield of $\eta_c$ and $J/\psi$ data were reconciled,
the corresponding LDMEs still could not solve the $J/\psi$ polarization puzzle.
Interestingly, Refs.~\cite{Chao:2012iv, Bodwin:2014gia} ommitted the feed down contributions from $\psi'$ and $\chi_c$ to the $J/\psi$ yield,
and claimed that they could explain the $J/\psi$ polarization data,
however, Refs.~\cite{Gong:2012ug, Shao:2014yta, Han:2014jya, Zhang:2014ybe} indicated that,
the inclusion of the feed-down parts would ruin the results.
In sum, the $J/\psi$ polarization puzzle is still the most challenging question in high energy physics waiting for new explorations.
The mess of the situation can actually be attributed to the difficulty in the determination of the LDMEs.
As is going to be seen later in this paper, ordinary fitting procedure is incapable of this question.
It would be urgent to break through and bring out a practical strategy,
which is able to definitely either solve the $J/\psi$ polarization puzzle or phenomelogically disprove NRQCD.


{\it Criticism of the extant fitting strategies.}---
Before putting forward the approach, we first outline the procedure of determining the LDMEs on the market.
Here we focus on the direct $\psi$ ($J/\psi$, $\psi'$) production case, in which the cross section can be expressed as~\cite{Bodwin:1994jh}
$d\sigma(\psi)=\sum_ndf_n\langle O^{\psi}(n)\rangle$,
where $f_n$ is the short-distance coefficient (SDC) for producing a $c\bar{c}$ pair with quantum number $n$,
and $\langle O^{\psi}(n)\rangle$ is the corresponding LDME.
Notice that NRQCD is an effective theory,
we may expect its predictions to have an intrinsic deviation (which might not be very large, however, does exist) from the reality.
In addition, our concerns are always limited to specific processes
(sometimes because of the lack of knowledge on other processes,
which is due to e.g. experiment is lacking or higher-order corrections are large).
For this reason, we regard two sets of the LDMEs leading to close predictions in the processes we concern as "equivalent for these processes".
Further progress in both theoretical calculation and experimental measurement would distinguish the "equivalent" sets of the LDMEs.
Up to QCD NLO, perhaps the $\psi$ hadroproduction is the only process in which the dominant contributions are all counted.

When we fit the $\psi$ yield data, the standard deviation ($\overline{\chi^2}$), which is defined as
\be
\overline{\chi^2}=\frac{1}{D}\sum_d(\frac{\sigma^{th}_d-\sigma^{ex}_d}{\epsilon_d})^2, \label{eqn:chi2}
\ee
is a quadratic function of the LDMEs.
Here, $\sigma^{th}_d$, $\sigma^{ex}_d$ and $\epsilon_d$ denote the theoretical prediction,
and the experimental central value and error for the $d$th experimental data point, respetively,
and $D$ is the degree of freedom in the fit.
By way of illustration, we only take the three color-octet (CO) matrix elements,
$\langle O^{\psi}(^1S_0^{[8]})\rangle$, $\langle O^{\psi}(^3S_1^{[8]})\rangle$ and $\langle O^{\psi}(^3P_0^{[8]})\rangle$,
as to be determined.
To keep the homogeneity of the dimensions of the CO LDMEs, in this paper,
we definde $f_{^3P_J^{[8]}}$ and $\langle O^{\psi}(^3P_0^{[8]})\rangle$ by multiplying and dividing by a factor of $m_c^2$, respectively.
For convenience, $\langle O^{\psi}(n)\rangle$ is alternatively abbreviated to ${\cal O}^{\psi}_n \times 10^{-2}\gev^3$ in the following,
with $n=$1,2,3 representing $^1S_0^{[8]}$, $^3S_1^{[8]}$ and $^3P_0^{[8]}$, respectively.

Ordinary fitting procedure is to solve the equations,
$\partial\overline{\chi^2}/\partial{\cal O}_n=0$,
to fix the values of the LDMEs at which the $\overline{\chi^2}$ reaches its minimum.
However, Ref.~\cite{Ma:2010yw} found that the SDCs for the three CO channels roughly satisfy a linear relation
\be
f_{^3P_J^{[8]}}=r_0f_{^1S_0^{[8]}}+r_1f_{^3S_1^{[8]}}, \label{eqn:sdclinear}
\ee
thus, only two of the three LDMEs can be fixed through the fit of the yield data.
For instance, the cross section for direct $\psi$ hadroproduction can be expressed as
\be
d\sigma(\psi)=f_{^1S_0^{[8]}}M_0^{\psi}+f_{^3S_1^{[8]}}M_1^{\psi}, \label{eqn:fm}
\ee
where $M_0^{\psi}$ and $M_1^{\psi}$ are defined by
\bea
M_0^{\psi}&=&\langle O^{\psi}(^1S_0^{[8]})\rangle+r_0\langle O^{\psi}(^3P_0^{[8]})\rangle, \NO \\
M_1^{\psi}&=&\langle O^{\psi}(^3S_1^{[8]})\rangle+r_1\langle O^{\psi}(^3P_0^{[8]})\rangle. \label{eqn:m0m1}
\eea
One can fit the yield data and obtain the values of $M_0$ and $M_1$ by employing Eq(\ref{eqn:fm}).

The reduction strategy provided in Eq.(\ref{eqn:sdclinear}) is feasible to work on the $\psi$ yield,
nevertheless, we find that it is not suitable for the polarization problem.
On the one hand, $\lambda$ is sensitive to as many as two parameters, namely $M_1$ and $r_1$;
even a slight variation of the two parameters can cause dramatic change of $\lambda$.
On the other hand, Eq.(\ref{eqn:sdclinear}) is only an approximate relation;
$r_0$ and $r_1$ are different in different kinematic regions.
For instance, for the CDF experimental condition,
in the range $7\gev<p_t<30\gev$, $r_0=3.9$ and $r_1=-0.56$,
while in the range $11\gev<p_t<30\gev$, $r_0=3.5$ and $r_1=-0.53$.
The difference of $r_1$ for the two $p_t$ ranges is large enough to completely change the predictions of the polarization.

{\it New methodology and its implementation.}---
Ref.~\cite{Zhang:2014ybe} provided an evidence for the VSR,
which is the most fundatmental bases of NRQCD (otherwise, the infinite higher excited Fock states of $c\bar{c}$ will be involved).
To this end, we constrain our discussions within the extent where this rule is not violated.
Under this assumption, $^1S_0^{[8]}$ channel cannot dominate the $\psi$ production,
thus, the $^3S_1^{[8]}$ and $^3P_J^{[8]}$ channels must contribute a large part and the cancellation between them would be significant.
We emphasize that, the cancellation is actually natural,
since, at QCD NLO or higher order, the two channels are associated;
only the combination of the two channels is divergence free and NRQCD scale independent.
We need only to argue that for any process, the cancellation does not cause unphysical results (negative cross sections).
Actually, it is a direct conclusion of the fragmentation mechanism.
The combination (as well as the cancellation) of the two channels can be carried out at fragmentation-function level.
In addition, since the $^1S_0^{[8]}$ channels cannot saturate the $\psi$ hadroproduction,
the leading and next-to-leading power~\cite{Ma:2013yla, Ma:2014eja} terms in the combination of the $^3S_1^{[8]}$ and $^3P_J^{[8]}$ channels must be positive.
Consequently, the combination of the fragmentation functions (FFs) multiplied by the LDMEs of the two channels must be positive definite.
So what we need to work out is how much do they cancel,
or equivalently, what is the value of $R_\psi\equiv\langle O^\psi(^3S_1^{[8]})\rangle/\langle O^\psi(^3P_0^{[8]})\rangle$.

We will find that the $\psi$ polarization is extremely sensitive to $R_\psi$ while the $\psi$ yield is not.
This brings us to a subtle circumstance that small variations of $R_\psi$ result in equivalent LDMEs for the $\psi$ production,
however, give totally different predictions of the polarizations.
This is the exact reason why one cannot succeed in the explanation of the $\psi$ polarization
by employing the LDMEs obtained in the fit of the yield data by minimizing the $\overline{\chi^2}$.
Since for the $J/\psi$, the $^1S_0^{[8]}$ LDME obtained in Ref.~\cite{Zhang:2014ybe} has a large uncertainty,
while for the $\psi'$, it is totally unknown,
in our strategy, we assign it different values, and directly fit $\langle O^\psi(^3S_1^{[8]})\rangle$ and $\langle O^\psi(^3P_0^{[8]})\rangle$,
and find that the variation of $\langle O^\psi(^1S_0^{[8]})\rangle$ only leads to equivalent LDMEs for both $\psi$ yield and polarization.
Having this, we can assign it any possible value under the constraint of the VSR and Ref.~\cite{Zhang:2014ybe}.

In low and high $p_t$ region and forward (backward) rapidity ($y$) region,
large logs ($log(m_c^2/p_t^2)$, $log(p_t/E_\psi)$) are important.
We constrain our concerns in the kinematic region, $|y|<1.6$ and $7\gev<p_t<30\gev$ for $J/\psi$ and $11\gev<p_t<30\gev$ for $\psi'$~\cite{Shao:2014yta},
in order to keep the perturbative expansions safe.
For the yield, all the data in these region provided in Refs.~\cite{Acosta:2004yw, Aaltonen:2009dm, Aad:2011sp, Aad:2014fpa, Chatrchyan:2011kc} are included in our fit.
Note that the $\psi'$ polarization data in Refs.~\cite{Abulencia:2007us, Chatrchyan:2013cla} have large errors,
we would start from the study of the $J/\psi$, where, for the yield, the contributions from $\psi'$ feed down are counted,
while for the polarization, they are neglected,
under the consideration that $\psi'$ feed down only contribute a fraction of less than 10\% to the prompt $J/\psi$ production,
which cannot affect the $J/\psi$ polarization.
After acquiring some educational knowledge from the $J/\psi$ case, we would come back to study the $\psi'$ meson.

For the $\chi_c$ production, we adopt the same parameter choices as in Ref.~\cite{Jia:2014jfa}.
To calculate its contributions to the polarized $J/\psi$, we follow the scheme developed in Ref.~\cite{Gong:2012ug}.
The parameters adopted in the analysis of $J/\psi$ and $\psi'$ yield and polarization are identical to those adopted and obtained in Ref.~\cite{Zhang:2014ybe},
except for the values of $\langle O^{\psi}(^3S_1^{[8]})\rangle$ and $\langle O^{\psi}(^3P_0^{[8]})\rangle$.
Since the color-singlet (CS) LDME and $\langle O^{\psi}(^1S_0^{[8]})\rangle$ are identical to those in Ref.~\cite{Zhang:2014ybe},
we can expect that they are naturally consistent with the $\eta_c$ hadroproduction data~\cite{Aaij:2014bga}.

\begin{table}[htbp]
\begin{center}
\begin{tabular}{|c|c|c|c|c|c|c|}
\hline
${\cal O}^{J/\psi}_3$&1.50&1.60&1.70&1.80&1.90&2.00\\
\hline
${\cal O}^{J/\psi}_2$&0.898&0.934&0.971&1.008&1.044&1.081\\
\hline
$R_{J/\psi}$&0.599&0.584&0.571&0.56&0.549&0.540\\
\hline
$\overline{\chi^2}$&2.16&2.03&1.98&2.00&2.10&2.27\\
\hline
\end{tabular}
\caption{
The value of ${\cal O}^{J/\psi}_2$ and the corresponding $R_{J/\psi}$ and $\overline{\chi^2}$ at each specific value of ${\cal O}^{J/\psi}_3$.
The global error of ${\cal O}^{J/\psi}_2$ is $\pm 0.011$.
}
\label{table:jpsi}
\end{center}
\end{table}

We first directly fit the $\psi'$ yield and obtain ${\cal O}^{\psi'}_2=0.48\pm 0.02$ and ${\cal O}^{\psi'}_3=0.80\pm 0.05$,
where ${\cal O}^{\psi'}_1=0$ is set as default.
Employing them, associted with the $\chi_c$ predictions ,we can extract the direct part from the prompt $J/\psi$ yield data,
and directly fit the data and obtain ${\cal O}^{J/\psi}_2=1.0\pm 0.1$ and ${\cal O}^{J/\psi}_3=1.7\pm 0.1$,
which are consistent with our previous work~\cite{Zhang:2014ybe},
with $\overline{\chi^2}\approx 1.98$.
We remember that small deviations from the optimized values of the LDMEs provide equally good descriptions of the yield data,
for this reason, we fit ${\cal O}^{J/\psi}_2$ at each specific value of ${\cal O}^{J/\psi}_3$.
The results are listed in TAB.\ref{table:jpsi}.
Regarding Eq.(\ref{eqn:chi2}), the deviation of the yield curve for ${\cal O}^{J/\psi}_3=2.0$ from that for ${\cal O}^{J/\psi}_3=1.7$
is less than $10\%$ of the experimental error in average.
Higher order corrections, large log resummations, experimental errors,
or even numerical uncertainties and the intrinsic errors of an effective theory are comparable with that.
In other words, the LDMEs listed in TAB.\ref{table:jpsi} are equivalent for the $J/\psi$ yield.
We can summarize the LDMEs in TAB.\ref{table:jpsi} in a compact form as
\be
{\cal O}^\psi_2=k_\psi{\cal O}^\psi_3+b_\psi, \label{eqn:kb}
\ee
while for $J/\psi$,
\be
k_{J/\psi}=0.367,~~~~b_{J/\psi}=0.348\pm 0.011. \label{eqn:kbjpsi}
\ee

Employing Eq.(\ref{eqn:kb}), we can fit the $J/\psi$ polarization data.
The degree of freedom has been reduced to one.
Since the measurement of CDF Run I and Run II are contradict with each other, we give up using the Run I data for their large uncertainties.
Considering that, the polarization is a ratio; even a slight error (as small as 20\%) can cause significant deviation,
we drop the data in low $p_t$ region, where the precision provided by the perturbative expansion is quite difficult to control.
Only the $p_t>10\gev$ data in Refs.~\cite{Abulencia:2007us, Chatrchyan:2013cla} are adopted in our fit.
Including the contributions from $\chi_c$ feed down and excluding those from $\psi'$,
we obtain the value of $R_{J/\psi}$ as
\be
R_{J/\psi}=0.546\pm 0.006. \label{eqn:K}
\ee
We emphasize again that $R_{J/\psi}$ is the only parameter to govern the $J/\psi$ polarization, as long as the VSR is kept.
For instance, if we fix $R_{J/\psi}$ and vary $\langle O^{J/\psi}(^1S_0^{[8]})\rangle$ from its upper to lower bound obtained in Ref.~\cite{Zhang:2014ybe},
or vary ${\cal O}^{J/\psi}_3$ from 1.5 to 2.0,
the corresponding change of $\lambda$ is less than 0.02 (most of the time, much smaller than this).
Accordingly, Eq.(\ref{eqn:kbjpsi}) and Eq.(\ref{eqn:K}) provide the uncorrelated form of the LDMEs;
the uncertainties of $b_{J/\psi}$ and $R_{J/\psi}$ describe those of the $J/\psi$ yield and polarization, respectively.

\begin{figure}
\center{
\includegraphics*[scale=0.3]{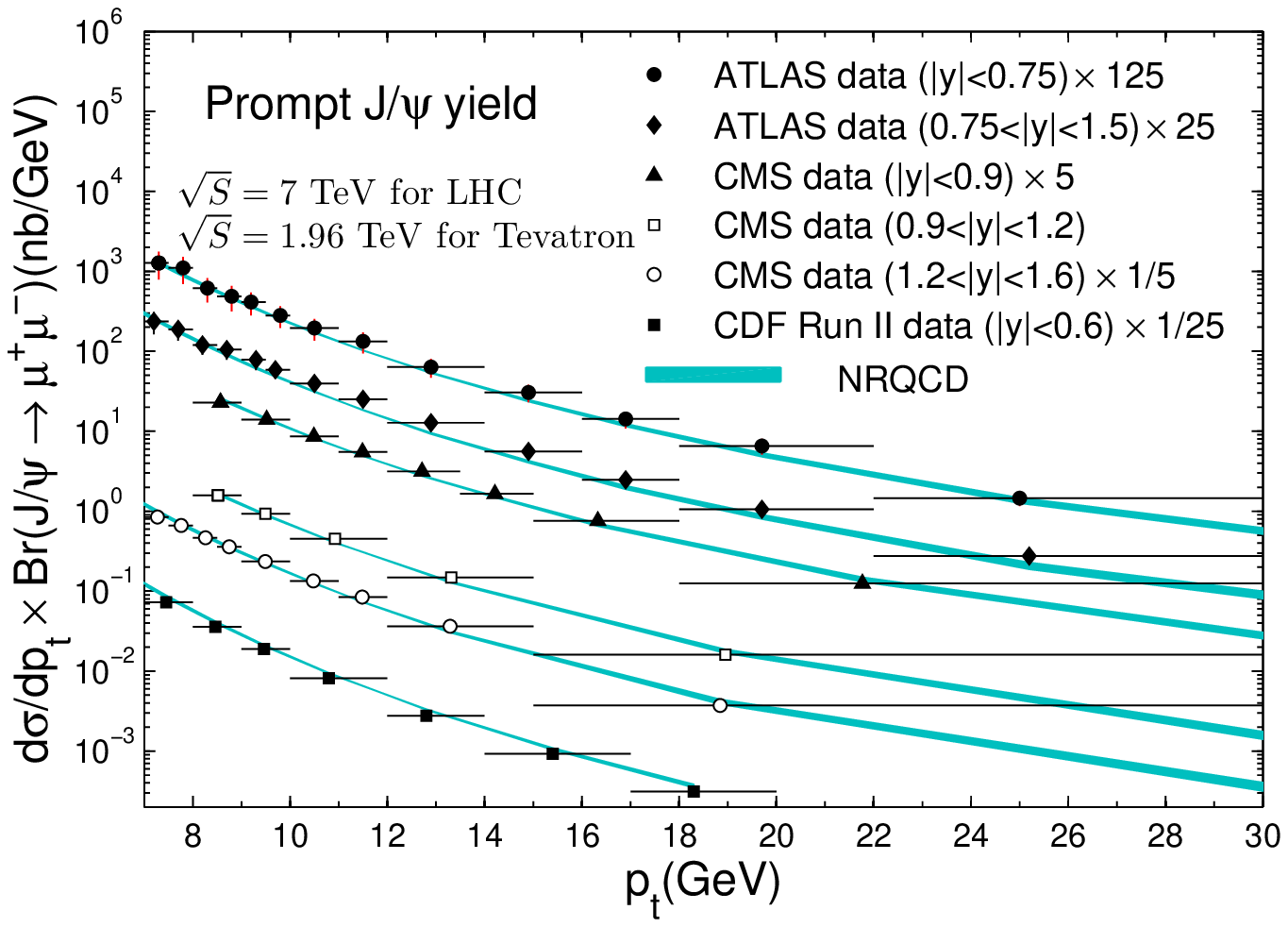}
\includegraphics*[scale=0.3]{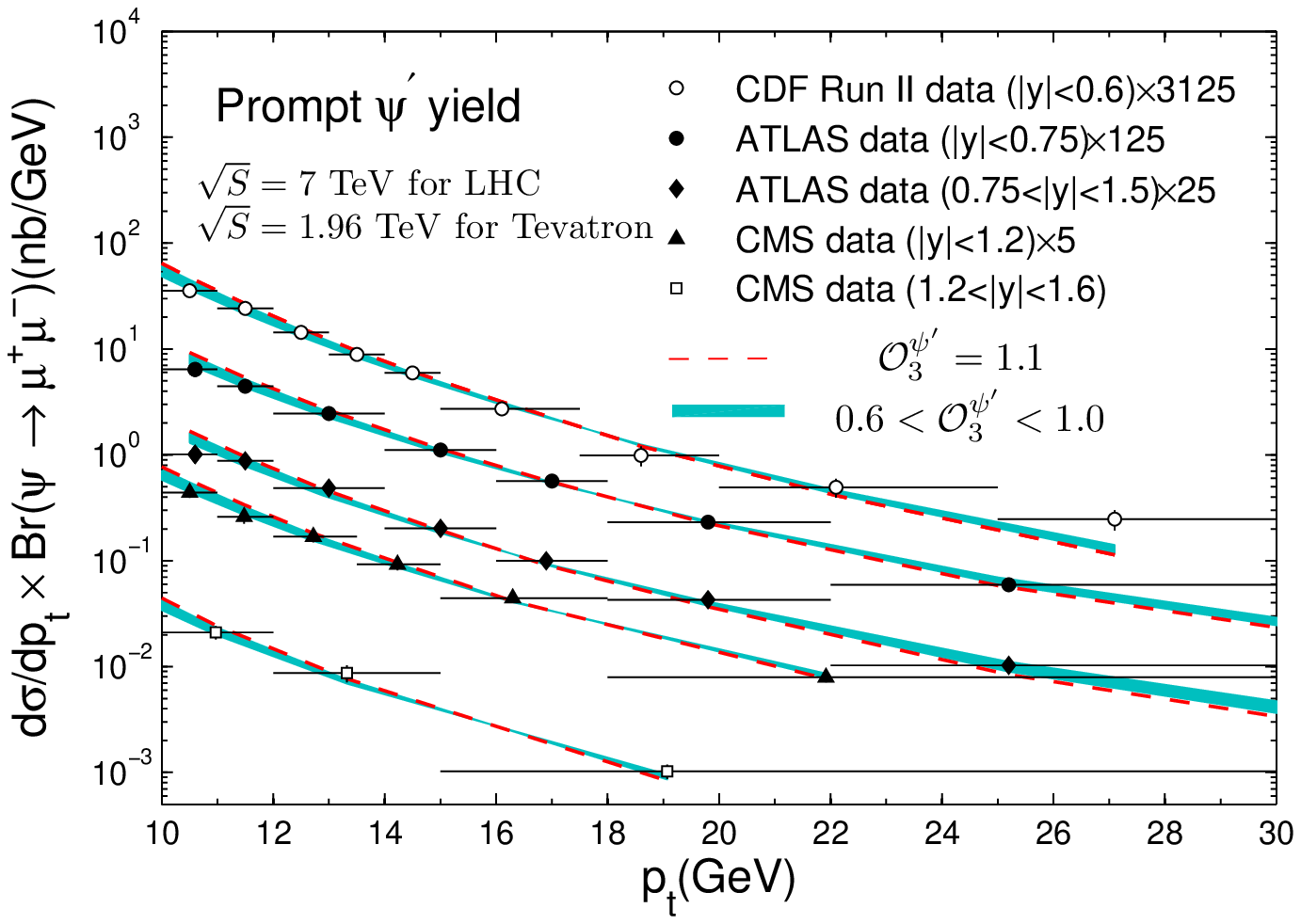}
\caption {\label{fig:yieldm}
$J/\psi$ and $\psi'$ yield at the Tevatron and the LHC in medium $p_t$ region.
All the LDMEs follow Eq.(\ref{eqn:kb}) in association with Eq.(\ref{eqn:kbjpsi}) for $J/\psi$ and Eq.(\ref{eqn:kbpsi2s}) for $\psi'$.
The data are taken from Refs.~\cite{Acosta:2004yw, Aaltonen:2009dm, Aad:2011sp, Aad:2014fpa, Chatrchyan:2011kc}
}}
\end{figure}

\begin{figure}
\center{
\includegraphics*[scale=0.3]{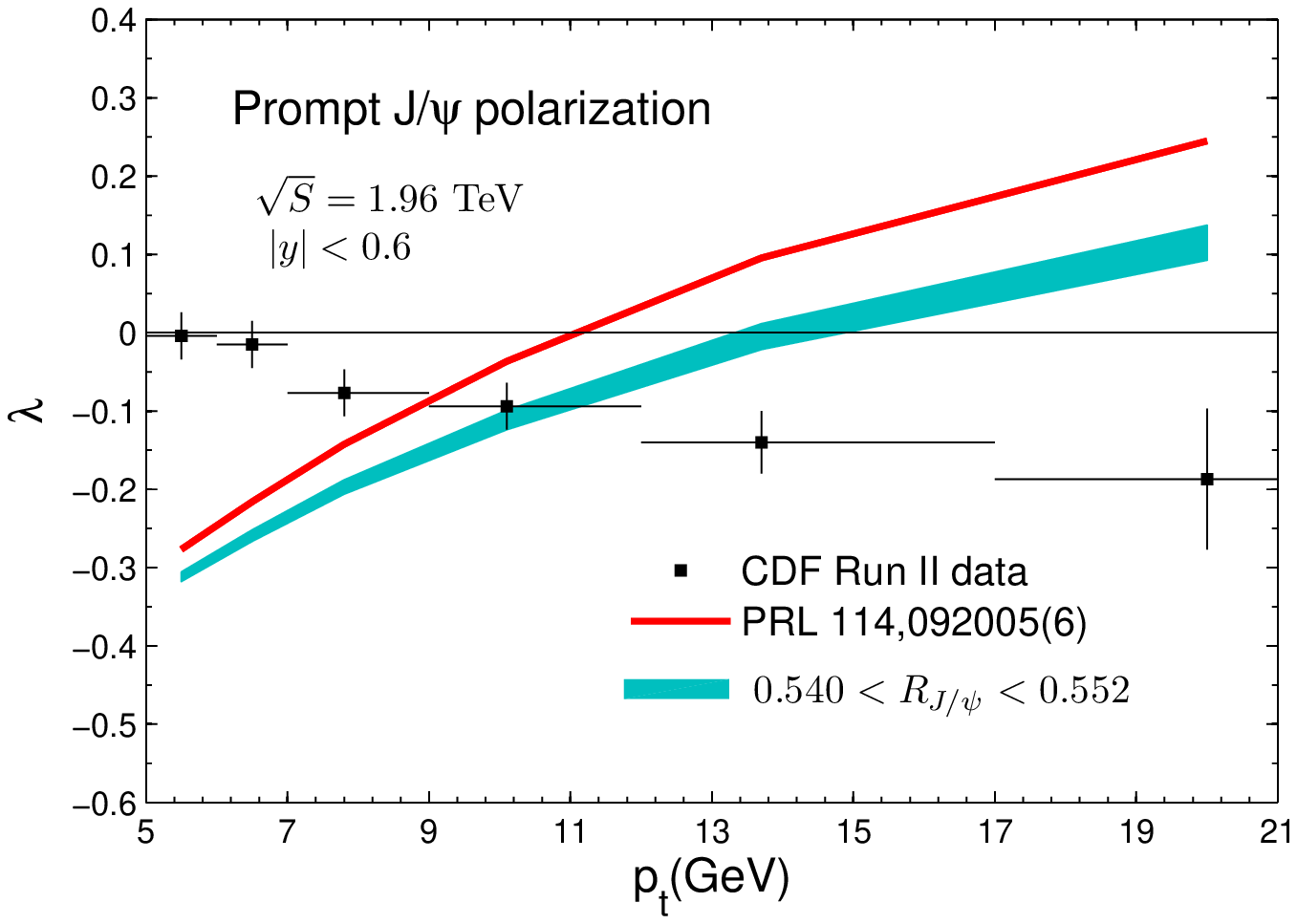}
\includegraphics*[scale=0.3]{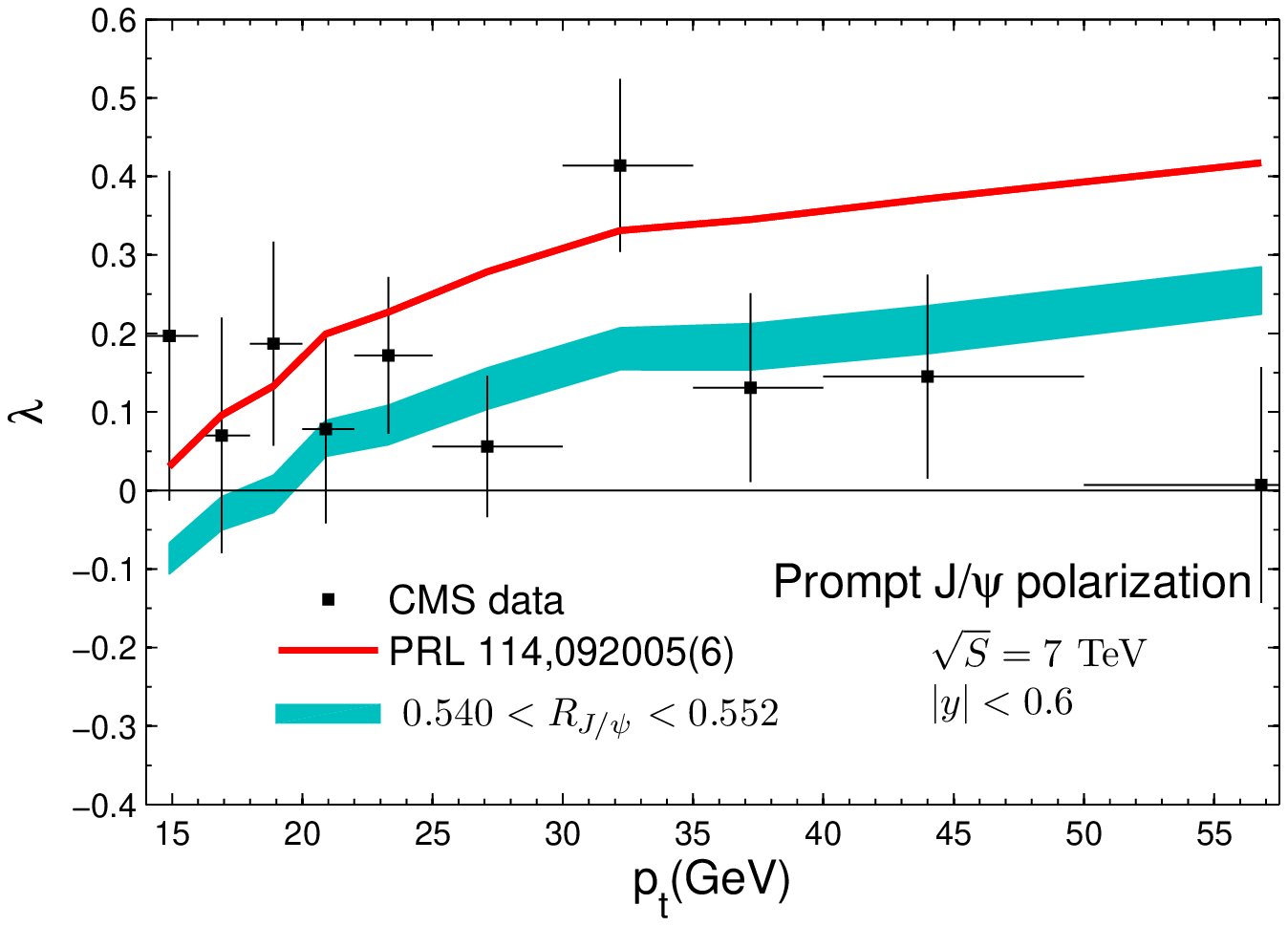}\\
\includegraphics*[scale=0.3]{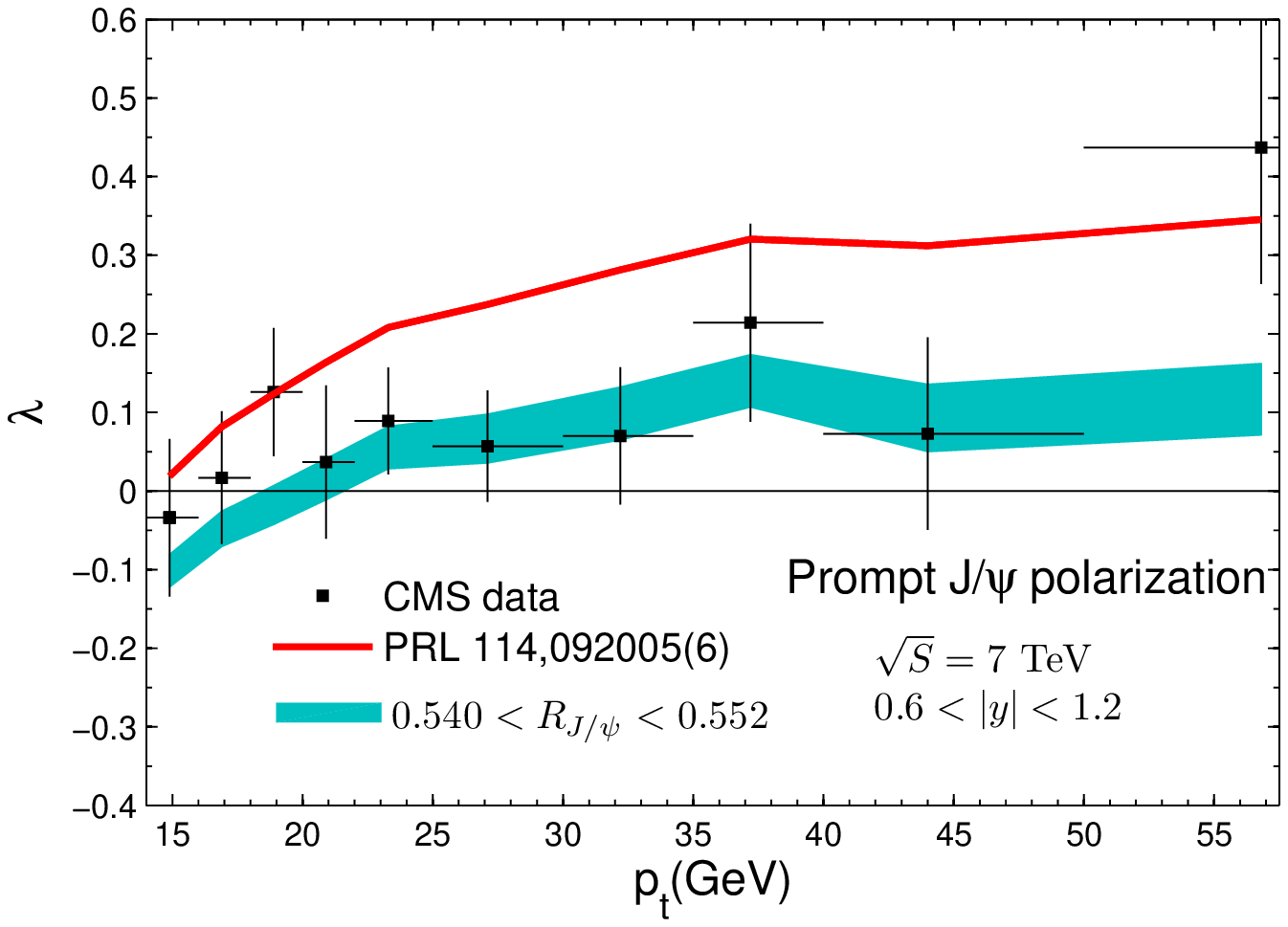}
\caption {\label{fig:poljpsi}
$J/\psi$ polarization at the Tevatron and the LHC.
The solid curves are produced with the LDMEs obtained in Ref.~\cite{Zhang:2014ybe},
while the bands are produced with the LDMEs corresponding to Eq.(\ref{eqn:K}).
The data are taken from Refs.~\cite{Abulencia:2007us, Chatrchyan:2013cla}.
}}
\end{figure}

The L.H.S. plot of Fig.\ref{fig:yieldm} displays the theoretical predictions versus data for the $J/\psi$ hadroproduction.
The bands are expanded by the curves for $1.7<{\cal O}^{J/\psi}_3<2.0$,
with the corresponding ${\cal O}^{J/\psi}_2$ obtained through Eq.(\ref{eqn:kb}) and Eq.(\ref{eqn:kbjpsi}).
The LDMEs in this range can provide equally good descriptions of the yield data.
However, as is shown in Fig.\ref{fig:poljpsi}, they result in totally different polarization predictions.
The solid curves are produced with the LDMEs obtained in Refs.~\cite{Han:2014jya, Zhang:2014ybe},
which corresponds to $O^{J/\psi}_3=1.7$ with Eq.(\ref{eqn:kb}) and Eq.(\ref{eqn:kbjpsi}) satisfied,
while the bands are produced with the LDME ranges obtained in Eq.(\ref{eqn:K}),
which corresponds to $1.88<{\cal O}^{J/\psi}_3<2.01$.
We can see that the CMS data are well described in our framework.
For the CDF data, the discrepancy between theory and experiment is larger.
We need to present the values of $\xi$ to acquire a better understanding of the problem.
For the $p_t=20\gev$ data, $\xi=1.2$ and 0.7 for the theoretical prediction and the experimental central value, respectively.
A correction of about only 20\% can fill this gap.
In the next part, we will further argue that, when $p_t<20\gev$,
the polarization is sensitive to the corrections while at higher $p_t$, it is not.
In this sense, we can say that,
the $J/\psi$ yield and polarization data as well as the $\eta_c$ yield data are reconciled within the NRQCD framework.

\begin{table}[htbp]
\begin{center}
\begin{tabular}{|c|c|c|c|c|c|c|c|}
\hline
${\cal O}^{\psi '}_3$&0.50&0.60&0.70&0.80&0.90&1.00&1.10\\
\hline
${\cal O}^{\psi '}_2$&0.353&0.395&0.437&0.479&0.521&0.563&0.605\\
\hline
$R_{\psi'}$&0.706&0.658&0.624&0.599&0.579&0.563&0.55\\
\hline
$\overline{\chi^2}$&1.23&0.79&0.53&0.44&0.53&0.79&1.22\\
\hline
\end{tabular}
\caption{
The value of ${\cal O}^{\psi '}_2$ and the corresponding $R_{\psi'}$ and $\overline{\chi^2}$ at each specific value of ${\cal O}^{\psi '}_3$,
while ${\cal O}^{\psi '}_1=0$ is set as default.
The global error of ${\cal O}^{\psi '}_2$ is $\pm 0.003$.
}
\label{table:psi2s}
\end{center}
\end{table}

\begin{figure}
\center{
\includegraphics*[scale=0.3]{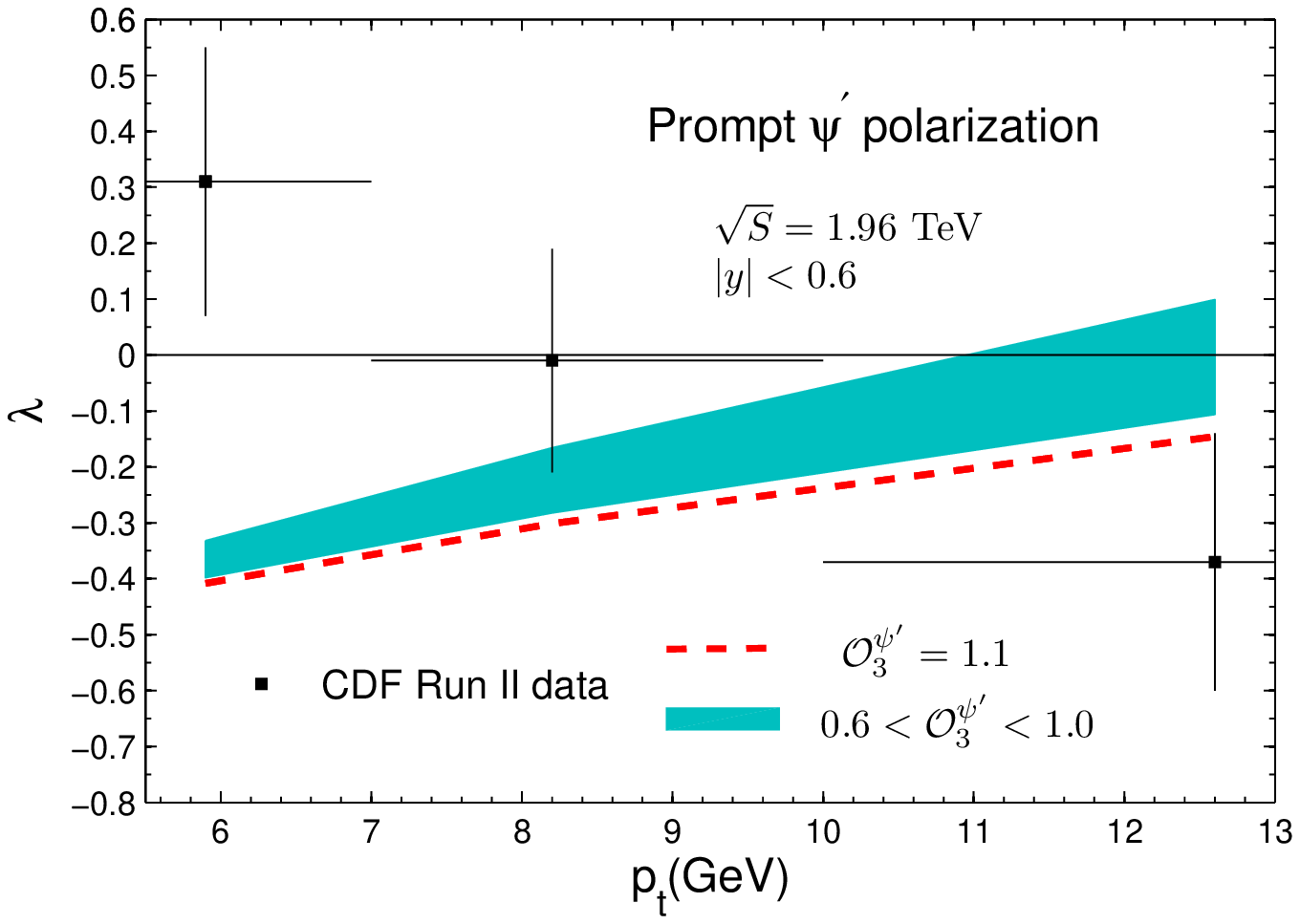}
\includegraphics*[scale=0.3]{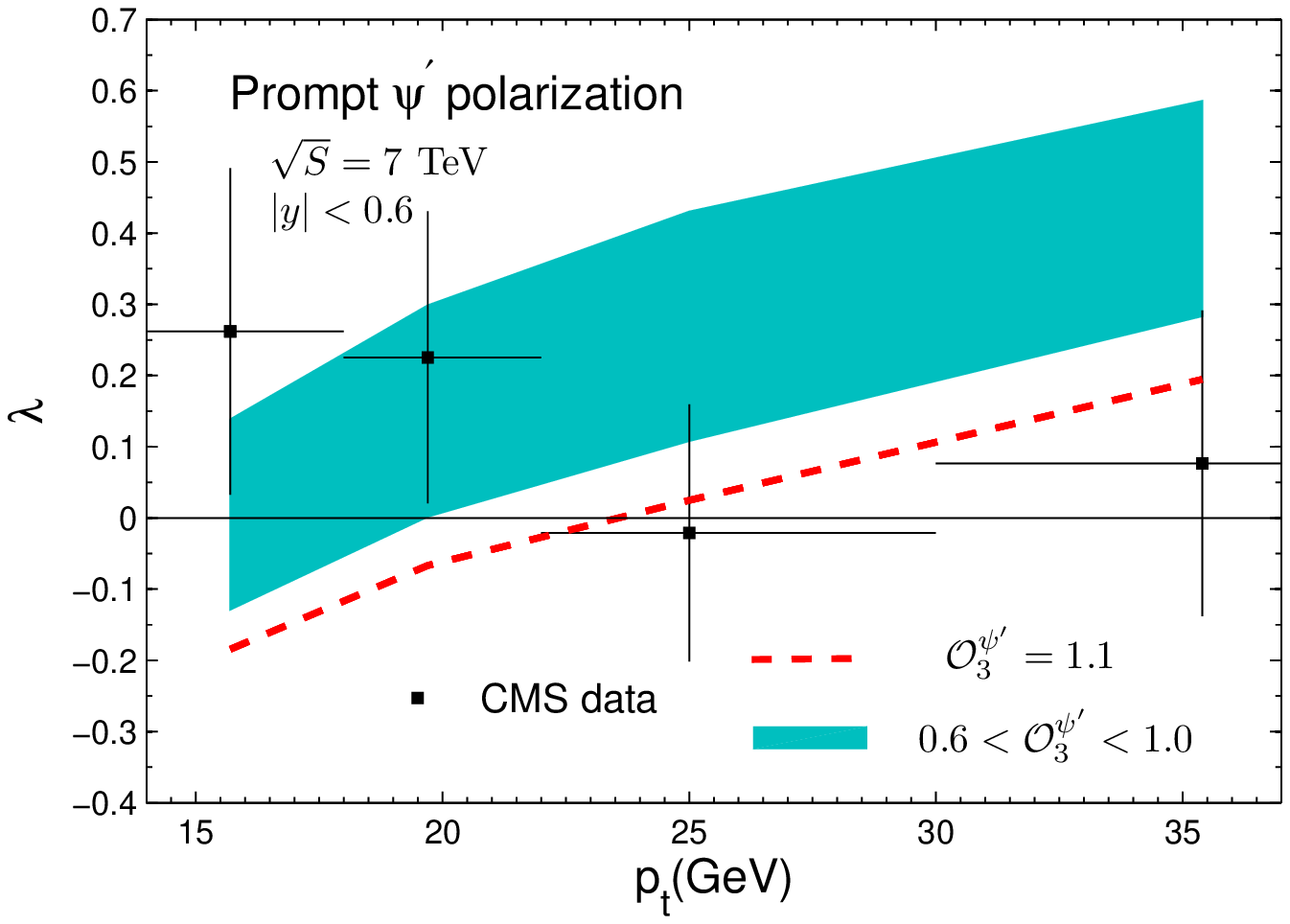}\\
\includegraphics*[scale=0.3]{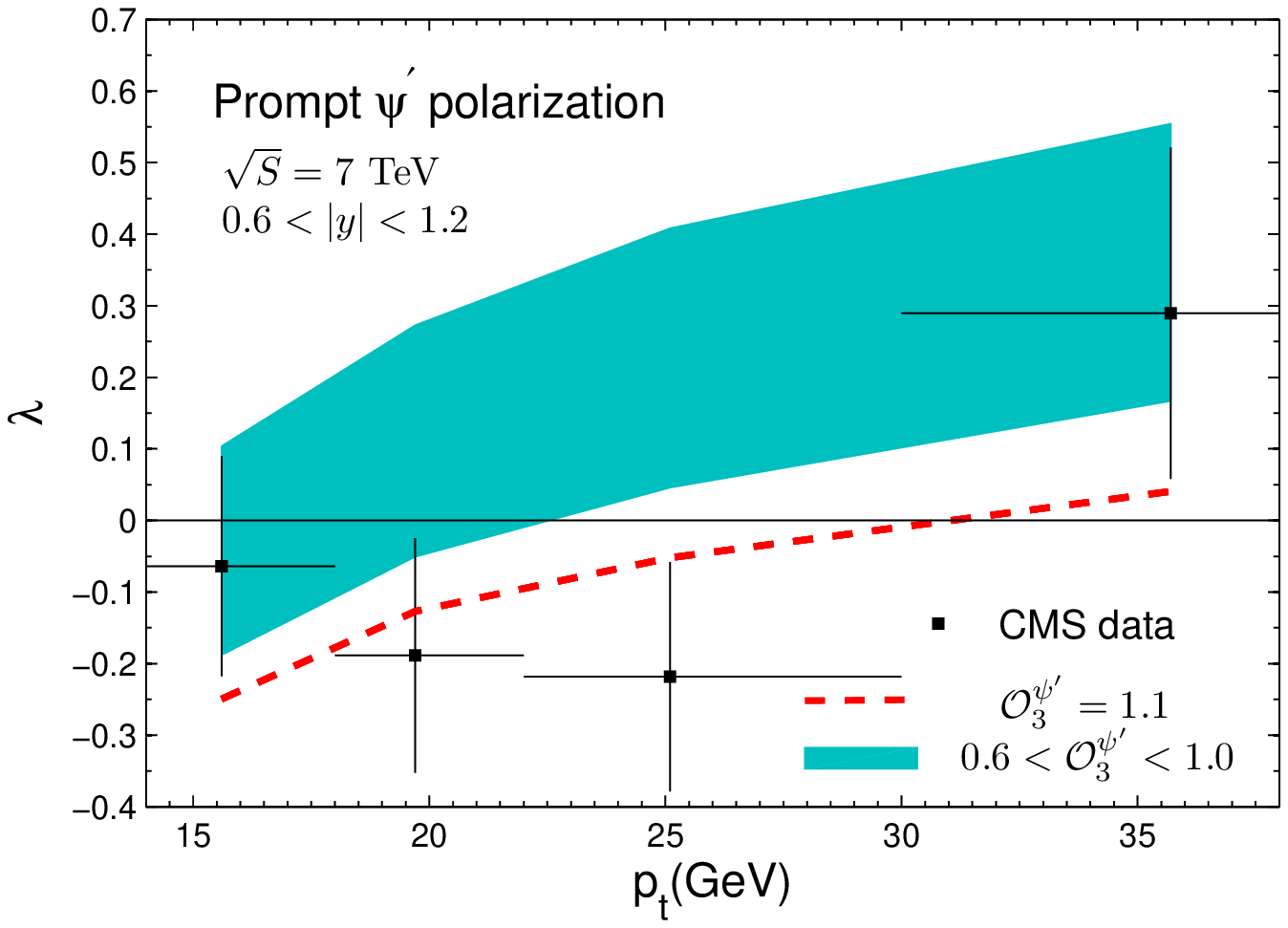}
\includegraphics*[scale=0.3]{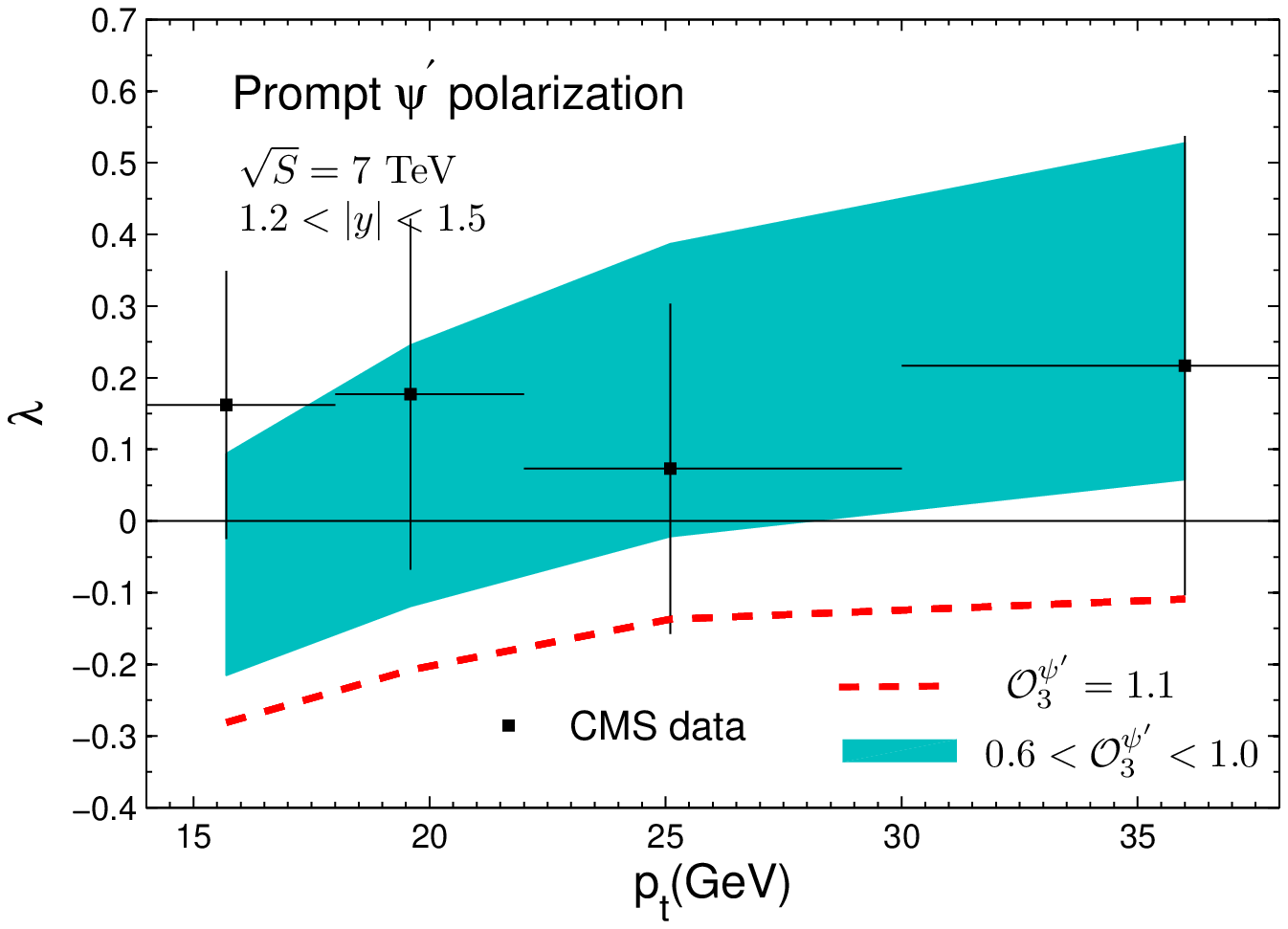}
\caption {\label{fig:polpsi2s}
$\psi'$ polarization at the Tevatron and the LHC.
All the LDMEs follow Eq.(\ref{eqn:kb}) and Eq.(\ref{eqn:kbpsi2s}).
The data are taken from Refs.~\cite{Abulencia:2007us, Chatrchyan:2013cla}.
}}
\end{figure}

Note that the CMS data for the $\psi'$ polarization is not monotonic with respect to the rapidity~\cite{Chatrchyan:2013cla},
we can expect that precise measurement will significantly change the central values.
For this reason, we would see how much our prediction of the polarization can reach,
while the yield data is well described,
instead of carrying out a fit of the experiment.
We can obtain a similar table, TAB.\ref{table:psi2s}, as TAB.\ref{table:jpsi} for $J/\psi$,
and the corresponding linear relation with
\be
k_{\psi'}=0.42,~~~~b_{\psi'}=0.143\pm 0.003. \label{eqn:kbpsi2s}
\ee
The bands in the R.H.S. plot of Fig.\ref{fig:yieldm} and in Fig.\ref{fig:polpsi2s} correspond to the range $0.6<{\cal O}^{\psi'}_3<1.0$,
in which, as is displayed in TAB.\ref{table:psi2s}, the $\overline{\chi^2}'s$ are quite small.
This range of the LDMEs lead to quite large polarization bands,
yet, not large enough to cover all the data points, as is shown in Fig.\ref{fig:polpsi2s}.
Therefore, we also present the curves for ${\cal O}^{\psi'}_3=1.1$,
which can cover the upper bound of the error bands of the CMS data in the rapidity range $0.6<|y|<1.2$.
The curves for the $\psi'$ yield also are in good agreement with the data,
albeit the $\overline{\chi^2}$'s are relatively larger.
Fig.\ref{fig:polpsi2s} and Fig.\ref{fig:yieldm} clearly manifest the fact that,
when Eq.(\ref{eqn:kb}) is held, the yield data can be well reproduced in very large range of the LDMEs,
while the polarization is extremely sensitive to $R_\psi$.
This is to say, it is almost impossible to describe the polarization data using the LDMEs obtained through the fit of the yield data;
even the variation of the yield curve is as slight as the intrinsic error of an effective theory,
the polarization will change dramatically.
We can also conclude that, it is almost certain that, when higher-precision polarization data come out,
the yield and polarization of $\psi'$ can both be well reproduced.

Having got ${\cal O}^{\psi '}_2$ and ${\cal O}^{\psi '}_3$ for ${\cal O}^{\psi '}_1=0$,
we can attempt to assign ${\cal O}^{\psi '}_1$ a larger value consistent with the VSR.
When ${\cal O}^{\psi '}_1=1.0$, we obtain $k_{\psi'}=0.42$ and $b_{\psi'}=0.115\pm 0.002$.
Following the same procedure, we find that, the phenomenological results does not change,
which proves that varying ${\cal O}^{\psi '}_1$ and redoing the fitting procedure only leads to equivalent LDMEs.

%

{\it Universality.}---
Since the $^3S_1^{[8]}$ and $^3P_J^{[8]}$ channels suffer from large cancellations,
we need to further discuss the stability of the results, or equivalently,
to answer the question, whether a slight (or even large) correction can ruin the phenomenology,
or equivalently, whether the solution of the LDMEs keeping the polarization lies in the region where the $\overline{\chi^2}$ for the fit of the $\psi$ yield is not too large.
%

Most of the time, the corrections are proportional to the NLO results.
Without lost of generality, we assume that the unpolarized SDCs remain,
while only the ratio of the transverse part to the longitudinal one ($\xi_{^3S_1^{[8]}}$ and $\xi_{^3P_J^{[8]}}$) changes.
Under this assuption, the linear relations in Eq.(\ref{eqn:kb}) hold, so do TAB.\ref{table:jpsi} and TAB.\ref{table:psi2s}.
Only the relation between $\lambda$ and $R_\psi$ changes.
We recalculate the values of $\lambda$ for specific values of $\xi_{^3S_1^{[8]}}$ and $\xi_{^3P_J^{[8]}}$ as a function of $R_\psi$ numerically,
and find that, in large $p_t$ region, especially $p_t>20\gev$, to obtain the previous results of the polarizations,
the corresponding $\overline{\chi^2}$'s for the fit of the $\psi$ yield are also small.
For instance, if $\xi_{^3P_J^{[8]}}$ decreases by 30\% (which is quite a large correction),
to keep the polarization of the $J/\psi$ or $\psi'$ invariant,
$R_\psi$ need to increase 0.05 at the most.
For both $J/\psi$ and $\psi'$,
the $\overline{\chi^2}$ for the fit of the yield data at the shifted $R_\psi$ is tolerable regarding TAB.\ref{table:jpsi} and TAB.\ref{table:psi2s}.
However, in the region $p_t<20\gev$, $\lambda$ is more sensitive to the corrections.
Most of the time, the corrections would enhance the longitudinal fraction, so,
we can expect that, the CDF data for the $J/\psi$ polarization will be better described when further corrections are counted.

We need also to investigate whether high $p_t$ data can help to fix the LDMEs in the absence of the polarization data,
if the FFs~\cite{Ma:2013yla, Ma:2014eja} are employed.
Actually, both the $^3S_1^{[8]}$ and $^3P_J^{[8]}$ SDCs scale as $p_t^{-4}$ in large $p_t$ limit,
thereafter, they will always be tangled.
For instance, the ratio of the curve for $O^{\psi'}_2=0.395$ and $O^{\psi'}_3=0.60$
to the curve for $O^{\psi'}_2=0.605$ and $O^{\psi'}_3=1.10$ is about 1.5 in large $p_t$ limit in midrapidity region,
while the difference of the corresponding $\lambda$ between the two sets of the LDMEs can be as large as 0.7.
Therefore, high $p_t$ data might provide slight constraints of the LDMEs,
still, it is impossible to describe the polarization with the LDMEs obtained employing only the yield data.

The only remaining question we would discuss is that, whether a global fit, employing both the yield and polarization data,
can give the same LDMEs as presented in Eq.(\ref{eqn:kbjpsi}) and Eq.(\ref{eqn:K}).
First we set the LDME for $^1S_0^{[8]}$ as a free parameter,
and obtain ${\cal O}^{J/\psi}_1=3.0\pm 1.6$, ${\cal O}^{J/\psi}_2=0.7\pm 0.2$, and ${\cal O}^{J/\psi}_3=1.3\pm 0.4$,
which contradicts with the $\eta_c$ hadroproduction data.
So, we set ${\cal O}^{J/\psi}_1=0.78$ according with Ref.~\cite{Zhang:2014ybe}.
The global fit gives ${\cal O}^{J/\psi}_2=1.02\pm 0.03$ and ${\cal O}^{J/\psi}_3=1.84\pm 0.06$,
the corresponding value of $R_{J/\psi}$ of which is $R_{J/\psi}=0.554$,
which is above the upper bound of the range in Eq.(\ref{eqn:K}).
So, we can conclude that, a global fit, even including the polarization data,
still cannot tackle the $J/\psi$ polarization puzzle.

{\it Summary.}---
In this paper, we discovered the unique key parameter which governs the $\psi$ polarization,
namely $R_\psi\equiv\langle O^\psi(^3S_1^{[8]}\rangle/\langle O^\psi(^3P_0^{[8]}\rangle$,
and reconciled all the charmonia yield and polarization data in midrapidity region within NRQCD framework.
When $R_\psi$ is fixed, varying the LDMEs hardly changes the polarization if the VSR is not violated.
Besides, we found that the polarization is extremely sensitive to $R_\psi$ even under the constraint of the yield data.
Accordingly, it is almost impossible to explain the polarization with the LDMEs fixed by the yield data.
Through a brief analysis, we found that the cancellation between $^3S_1^{[8]}$ and $^3P_J^{[8]}$ is natural,
and further corrections might change the values of the LDMEs but not the phenomenological results.
A global fit, even including the polarization data, is incapable of the determination of the LDMEs.

We thank Yan-Qing Ma for helpful discussions.
This work is supported by the National Natural Science Foundation of China (Nos.~11405268).


\end{document}